\documentclass{article}

\usepackage[utf8]{inputenc} 
\usepackage[T1]{fontenc}    
\usepackage{hyperref}       
\usepackage{url}            
\usepackage{booktabs}       
\usepackage{amsfonts}       
\usepackage{nicefrac}       
\usepackage{microtype}      
\usepackage{xcolor}         
\usepackage{algpseudocode}  
\usepackage{amsmath}        
\usepackage{graphicx}       

\usepackage{caption}
\usepackage{subcaption}
\usepackage{wrapfig}

\usepackage[final]{neurips_2023_ml4ps}

\newcommand{\simtool}{\textit{DeepSurveySim}}

\title{\simtool: Simulation Software and Benchmark Challenges for Astronomical Observation Scheduling}

\author{%
    M. Voetberg \thanks{Corresponding Author} \\
    Fermi National Accelerator Laboratory \\
    Batavia, IL 60510 \\
    \texttt{maggiev@fnal.gov} \\
    \AND
    Brian Nord \\
    Fermi National Accelerator Laboratory\\
    Batavia, IL 60510\\
    Department of Astronomy and Astrophysics, University of Chicago\\ 
    Chicago, IL 60637\\
    Kavli Institute for Cosmological Physics, University of Chicago\\
    Chicago, IL 60637 \\
    \texttt{nord@fnal.gov}
}

\begin{document}

\maketitle

\begin{abstract}

Modern astronomical surveys have multiple competing scientific goals.
Optimizing the observation schedule for these goals presents significant computational and theoretical challenges, and state-of-the-art methods rely on expensive human inspection of simulated telescope schedules.
Automated methods, such as reinforcement learning, have recently been explored to accelerate scheduling.
However, there do not yet exist benchmark data sets or user-friendly software frameworks for testing and comparing these methods. 
We present \simtool\ -- a high-fidelity and flexible simulation tool for use in telescope scheduling.
\simtool\ provides methods for tracking and approximating sky conditions for a  set of observations from a user-supplied telescope configuration. 
We envision this tool being used to produce benchmark data sets and for evaluating the efficacy of ground-based telescope scheduling algorithms, particularly for machine learning algorithms that would suffer in efficacy if limited to real data for training.
We introduce three example survey configurations and related code implementations as benchmark problems that can be simulated with \simtool.
\footnote{Code repository and application examples can be found at our \href{https://github.com/deepskies/DeepSurveySim}{GitHub repository, github.com/deepskies/DeepSurveySim.}}

\end{abstract}

\section{Introduction}
Modern astronomy experiments typically have multiple competing scientific goals.
Each of these goals requires the observation of a different kind of astronomical object -- e.g., static galaxies, time-varying stars (like supernovae), and transients (like planets).
In an astronomical survey campaign that spans months to years, scheduling observations of objects typically requires competing scheduling priorities.
For example, faint galaxies require multiple exposures of the same locations on the sky to reduce noise.
Transients and variables, on the other hand require exposures at many different locations in the sky, which diverts telescope time away from the location for the galaxies.
In addition, ground-based observing campaigns must contend with atmospheric conditions, which degrade data and can interrupt planned observations; weather conditions like clouds can interrupt a schedule or change the kinds of observations possible. 
~\footnote{We include a glossary of terms used in this paper in Table ~\ref{tab:glossary} in the Appendix.}

Optimizing a schedule amidst these complex and competing observational goals typically necessitates a forward model (like a simulation) of the night sky, observing conditions, and telescope operations.
These simulations are then used by a human to manually design a schedule or within optimization algorithms like Reinforcement Learning (RL).
All methods still typically require human intervention, even if only to validate a schedule produced with an automated approach.
However, the methods are not widely benchmarked or compared in detail, and a user-friendly, well-packaged, open-source simulation software to generate benchmark simulated data sets does not yet exist.

We present \simtool, an open-source software package for simulating ground-based surveys.
We also present three standard challenge problems for the comparison of scheduling methods and algorithms. 
\simtool\ can be used to generate data sets for these benchmark challenges. These "challenge" benchmarks can be used for evaluating traditional scheduling methods.
However, more importantly, they can supply easily accessible data generation tools for comparatively data-greedy Machine Learning algorithms. 

This manuscript is organized as follows. 
In Section~\ref{sec:relatedwork}, we describe the current state of the art for simulators and open problems.
In Section~\ref{sec:methods}, we describe the simulator model.
In Section~\ref{sec:problems}, we present three challenge problems that can be used as benchmarks for scheduling algorithms.

\section{Related Work}
\label{sec:relatedwork}

Individual experiments and surveys -- like the Sloan Digital Sky Survey~\cite{Santana_2021}, the Dark Energy Survey~\cite{2019arXiv191206254N}, and the Rubin Observatory's Legacy Survey of Space and Time (LSST)~\cite{lsst} -- typically create codebases for internal usage only for generating simulations of schedules, developing scheduling algorithms, and assessing potential observing strategies.
LSST conducted an open challenge within the collaboration for different research groups to create survey strategies optimized for their science \cite{2021arXiv210912060V} \cite{2016AJ....151..172G} \cite{2023ApJS..264...22G} \cite{2018AJ....155..128M} \cite{2018arXiv181200514F} \cite{2018arXiv181202204O} \cite{2018arXiv181200937V}.

Short survey campaigns can use tools like AstroPlan \cite{2018AJ....155..128M} for scheduling.
However, campaigns for larger surveys require complex simulators and more automated methods.
Some of the earliest work goes back to the Hubble Space Telescope's work with SPIKE  \cite{johnston_automated_1987}, which adapted factory automation planning techniques for ground-based telescopes. 
It was later applied for greedy search algorithms on moment-to-moment scheduling \cite{greedy_hubble}. 
These methods were a step forward for automating scientific discovery, but largely inaccessible at the time due to the large amounts of computing and data required to use them, and their inability to adapt to interruptions. 
More advanced automated approaches employ adaptive scheduling\cite{AdaptiveScheduler2023}, Reinforcement Learning (RL) algorithms \cite[e.g.,]{naghib_framework_2019}, or semi-supervised graph neural networks \cite[e.g.,]{cranmer_unsupervised_2021}. 
All of these classes of techniques require an environment, in the real world or otherwise, to be trained. 
Training these algorithms on real-world data sets is infeasible because there isn't enough historical data, and that historical data doesn't provide enough information for future campaigns. 

In any case, simulators and scheduling algorithms developed by surveys are rarely cross-compared or reported on in detail.
In the development of \simtool, we draw inspiration from other tools for autonomous control in our design, such as the driving simulation, CARLA \cite{Dosovitskiy17} and the flight system integration tool, OnAIR \cite{onair}. 
\simtool\ is based on simulation tools originally developed for DES and LSST \cite{2019arXiv191206254N}.

\section{\simtool: Simulator for ground-based optical telescopes}
\label{sec:methods}
\begin{figure}[h]
    \centering

  \makebox[\textwidth][c]{\includegraphics[width=.85\textwidth]{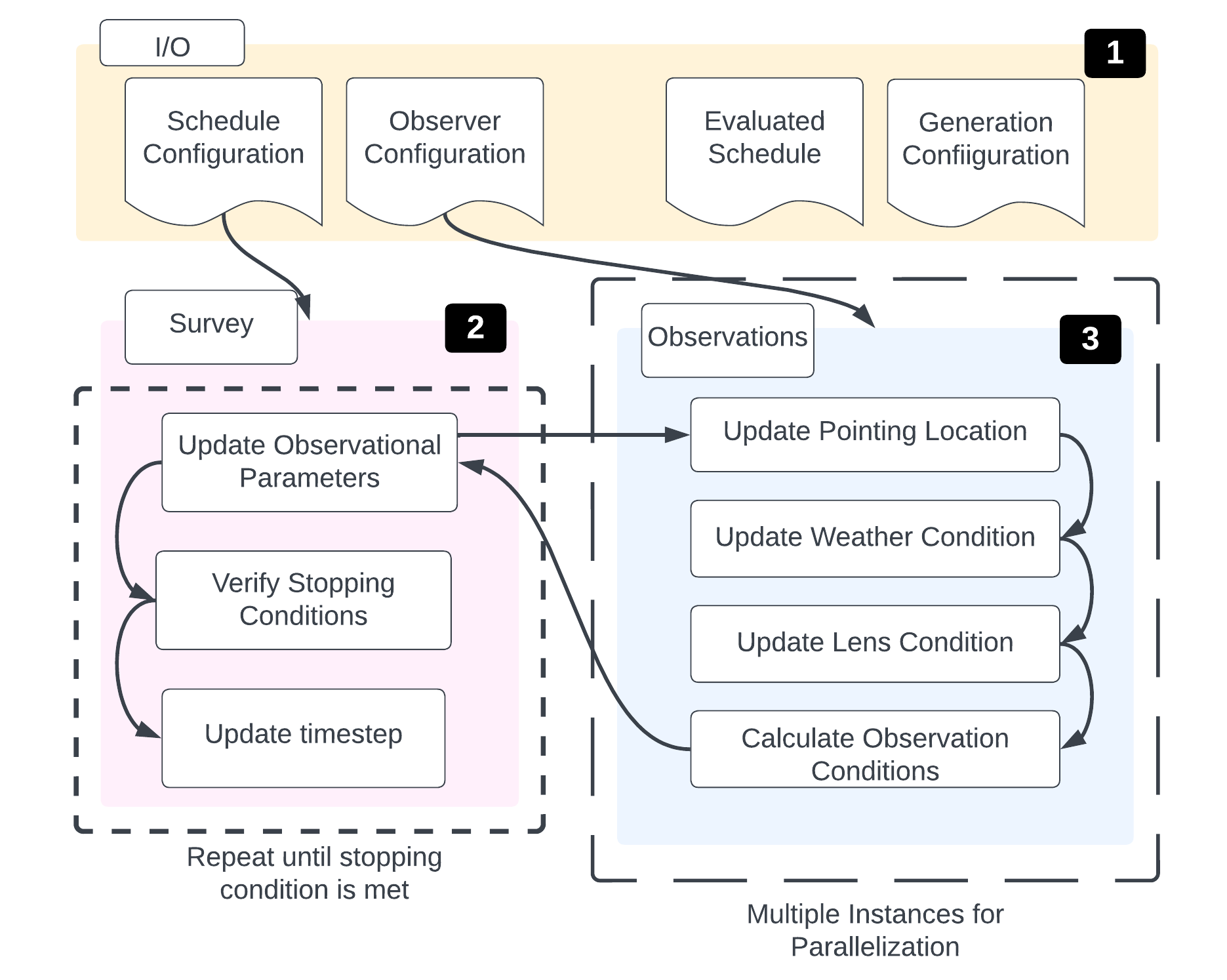}}
  \caption{Workflow diagram for \simtool. 
  The program initializes two different modules -- the survey (2, pink) and the observation variables (3, blue) via a configuration files (1, yellow). 
  The survey is then executed by calculating updated positions of sky locations and sky conditions until a user-defined stopping condition is met. }
    \label{fig:arch}
\end{figure}

We introduce \simtool, a light-weight and high-fidelity open-source simulation software to aid in scheduling observations for astronomical survey campaigns.
The simulator's calculations are deterministic regarding the positions of sky objects and the telescope's positions.
Calculations are performed on-demand and do not require large amounts of storage space to save generated schedules and related parameters. 
The simulator is designed to be highly flexible, and it comes with simple predefined configurations for particular observing goals and conditions. 
The simulator is primarily designed for use in multiple settings with Markov chain policy-style algorithms.
\simtool\ uses the \textit{Gymnasium} API \cite{towers_gymnasium_2023} as an inspiration for syntax. 
\simtool is specifically built to be used in conjunction with \textit{Gymnasium}, allowing for integration with many different Machine Learning frameworks that support the package. 
There are two main sections -- the `observation variables` and the `survey configuration` (Figure~\ref{fig:arch}). 
This allows the simulation to calculate only the variables required for the policy algorithm to advance and decrease overall compute time, or to add more variables and metrics for diagnostic purposes.
The user can designate the physical location of the observatory.
It also allows for separation between the type of schedule being generated and the observatory it is designed for, which is useful for testing method generalization.

The simulator has the following key features: 

\textbf{Independent Configuration Files} 
The simulator ingests configuration files, which are intended to be mixed and matched across varying observing scenarios (discussed in Section~\ref{sec:problems}) and observatories. 
This allows for high levels of reproducibility.

\textbf{Position Tracking} 
The simulator tracks four variables of the telescope's physical configuration:  the Right Ascension, Declination, Filter Wavelength Band, and Time (in Modified Julian Date).
These are specified by the user at each step or are ingested in bulk as an array.
The tracking functionality includes the option to account for time regarding the slew (telescope-repositioning), change of filters, or to include a level of inaccuracy in the repositioning system of the telescope, dependent on the angular distance between two observations. 

\textbf{Condition Tracking} 
The simulator uses the packages \textit{Astropy} \cite{astropy} and \textit{SkyBright} \cite{Neilsen:2019ump} for condition calculation. 
The simulator's calculation includes the position and brightness of the sun and moon relative to the telescope, approximated atmospheric conditions, and the position of the telescope relative to the earth. 
The full list of variables and their extended definitions can be found in Table~\ref{tab:vars} in the Appendix.
These allow a policy to walk through any given initial conditions and calculate a representation of the sky at a particular point in time, which permits the evaluation of the effectiveness of the scheduling algorithm.

\textbf{Approximate Weather} Given reference sky conditions for an observatory, the \simtool\ selects approximate seeing and cloud conditions based on the time of year. 
It takes as input the mean cloud cover for that time of year -- using reference weather station data supplied by a user, dependent on the location of their observatory -- and applies a threshold to determine if seeing and cloud extinction would be impacted.  

\section{Benchmark Challenge Problems} \label{sec:problems}
We provide a set of problems with corresponding code implementations. 
However, we encourage using these problems as a template to construct more specific scenarios to simulate.

\subsection{Problem 1: Maximizing Observation Quality in a Predetermined Set of Survey Fields}
\label{sec:problem_1}

Maximizing observation quality is one of the simplest problems to consider. 
Addressing this challenge typically requires producing a 'greedy' algorithm that can predict the next step that has the highest-quality observations, and produce the paired action for the telescope to move it to that site. 
We use the effective exposure time $\tau_{eff}$ defined as a simple proxy metric for ensuring the observation is high-quality and low-noise: 

\begin{equation}
    \tau_{eff} = (\eta * \frac{0.9}{\textsc{fwhm}})^2 * \frac{b_{ref}}{b} *\tau_{exposure};   
    \label{eq:teff}
\end{equation}

$\eta$ is the amount of light from a target reaching through the atmosphere as measured by instrumentation; $\textsc{fwhm}$ is the blurring of an image resulting from instrumentation and atmospheric conditions;  $\frac{b_{ref}}{b}$ is the magnitude of the sky brightness measured against a reference moonless night; and $\tau_{exposure}$ is the exposure time of a given observation. 
Further explanation can be found in \cite{neilsen_dark_2019}. 

This problem lacks realism because surveys generally have much more specific and complicated goals. 
Nevertheless, this example can be useful for validation of a given method.
This can be done by selected a small subset of sites (such as a small cluster along the equator), and verifying the method created the observation schedule for this site that provides the maximum schedule quality according to Equation~\ref{eq:discrete_p1}. 

A schedule that aims to maximize observation clarity for all images can be evaluated directly. 
The quality ($R$) of schedule $s_n$ over the discrete time interval $T = [t_0, t_{final}]$, where $t_0$ is the starting time of the schedule and $t_{final}$ is the end, is given by: $  R_{s_n} =  \frac{1}{||T||} \Sigma^{t_{n}}_{t=0} \tau_{eff_t} $.

Consider the scenario in which there is a finite number of pre-designated observation sites ($S$) and in which the observation schedule is created on the fly.
Then, the overall survey quality can be measured with 

\begin{equation}
        R_{s_n} =  \frac{1}{||T||}\frac{\Sigma^{t_{n}}_{t=0} \tau_{eff_t}(s_{selected_t})}{\Sigma^{t_{n}}_{t=0} max \{\tau_{eff_t}(s_t):s \in S \} }.
        \label{eq:discrete_p1}
\end{equation}

where $s_{selected_t}$ is the site selected for the observation at time $t$. 

This means that each step of the scheduling algorithm only requires the current state to make a decision about the next observation.
This produces a schedule that can be stopped and started to account either for unscheduled downtime.
We provide an example of a schedule created with this evaluation metric in Figure~\ref{fig:rl_solution} in the Appendix.

\subsection{Problem 2: Maximizing Observations of an Object with Low Visibility}

In some cases, there are objects of interest that are rarely visible -- e.g., transients that are in motion (like comets) or statistically rare events (like supernovae).
Transients have a deterministic trajectory, the observation of which can be explicitly accounted for in the environment as a rule for the schedule design: at time $t$, the action is always $a$ without being allowed to deviate.
The remainder of the schedule is planned around this constraint.
Rare events require deviations from the preset schedule to account for the chance of observing something in a given field.
One of the major challenges in this scenario is enabling the schedule to recover from the observation of the stochastic event.
This logic can also be applied in the case where weather events cause a stochastic variation in observability of a given part of the sky.
This presents an opportunity for the scheduling problem to be treated as a multi-objective optimization problem, which is represented by: 

\begin{equation}
    R_{s_{n}} = \frac{1}{||T||} (||\{s_i: s_i \in S_{interest}\}|| + \lambda \Sigma^{t_{n}}_{t=0}  
    \tau_{eff_t}(s_t)),
    \label{eq:multi_obj}
\end{equation}
where $S_{interest}$ denotes the condition of the observation containing the desired object, and $\lambda$ is a user-defined hyperparameter that determines the weight of observations  $ s \notin S_{interest}$. 

\subsection{Problem 3: Maximize Uniformity of Image Quality Across Sites}
 
Amongst the three example problems presented in this work, this problem is the most generic and has the widest application.  
A typical observing goal is uniformity of image quality $\theta$ across the survey.
This can be defined numerically as:

\begin{equation}
     R_{S_{n}} = \frac{1}{||T||* Var(\{\tau_{eff}(s_i): s_i \in S_n)\}}* \Sigma^{t_{n}}_{i=0} 
     \begin{cases} 
      \tau_{eff}(s_i) & \tau_{eff}(s_i) \geq \theta \\
      0 & \tau_{eff}(s_i) < \theta  \\
   \end{cases}.
    \label{eq:uniform_seeing}
\end{equation}

We must also prevent the algorithm from "cheating" and selecting only one location.
In our formulation, we determine that the schedule should be assigned a quality of 0 if it does not capture all sites at least $N$ times. 

\section{Conclusions}
Scheduling is not an easy task, and telescope scheduling, with all its possible stochastic interruptions, is even less so. 
We hope the introduction of a standardized simulation tool \simtool\ will make the evaluation of novel approaches more viable. 

\textbf{Limitations} 
This project is limited to deterministic variables, so random events in either atmosphere conditions or rare cosmological events are not modeled. 
The approximations of atmospheric conditions used by the simulation only holds for altitudes $\geq 20^{\circ}$ so observations very close to the horizon are inaccurate. 
Included weather condition calculations are based on historical measurements, so will not capture changes in the environmental trends. 

\begin{ack}
This work was produced by Fermi Research Alliance, LLC under Contract No. DE-AC02-07CH11359 with the U.S. Department of Energy, Office of Science, Office of High Energy Physics. Publisher acknowledges the U.S. Government license to provide public access under the DOE Public Access Plan DOE Public Access Plan. 

We acknowledge the Deep Skies Lab as a community of multi-domain experts and collaborators who’ve facilitated an environment of open discussion, idea-generation, and collaboration. This community was important for the development of this project. 

We thank Aleksandra \'Ciprijanovi\'c for her assistance in editing this manuscript, Franco Terranova  and Shohini Rhae for their assistance in testing the preliminary version of the package, and Eric Neilsen  Jr. for his domain expertise. 
\end{ack} 

\newpage

\section*{Appendix}

 \begin{table}[h]
    \centering
    \caption{Glossary of terms and associated symbols used throughout this paper, in order of appearance. 
    These are supplied for reference and ease of reading.}

     \begin{tabular}{p{0.3\linewidth} | p{0.1\linewidth}| p{0.5\linewidth}}
        Name & Symbol & Definition \\
        \hline 
        Schedule & -- & Sequence of observations ordered by time\\
        Survey & -- & Collection of observations, unordered \\ 
        Zenith Angle & $\phi$ & 90$^{\circ}$  tangential to the Earth's surface \\
        Markov Decision Process & MDP & Decision making process that takes the current state, increments it using a pre-defined policy, and calculates the benefit of this new state\\
        Reinforcement Learning & RL & MDP Based learning schema where a optimization algorithm learns the policy used to increment the state \\ 
        Right Ascension & RA &  Angular distance east or west of the point in the celestial equator where the sun is centered during the Spring equinox \\
        Declination & Decl. & Angular distance north or south of the celestial equator \\
        Modified Julian Date & MJD & Number of days since 17 November 1858\\
        Site/Observation/Pointing & -- & The location of a singular measurement from the telescope \\
        Discrete Time Intervals & $T$ & Collection of time steps a schedule covers\\
        Possible Sites & $S$ & Collection of observations included in a survey \\
        Schedule Quality & $R_{s_n}$ & The cumulative quality of all the observations in a schedule\\
        -- & $\lambda$ & Weighting factor used to scale observations that are not in the target survey \\
        Quality Threshold & $\theta$ & Lower threshold on the designated quality to determine if an observation is within the standards required for the survey \\
        Cardinality Threshold & $N$ & Lower threshold on the number times a site must be measured during a survey\\
     \end{tabular}
     \label{tab:glossary}
 \end{table}
 
 \begin{table}[h]
     \centering
     \caption{All possible variables included in the simulation framework. At the discretion of the user, if a variable is not directly involved in the evaluation of the schedule $R$, the variable can be removed to increase the speed of calculations.}

     \begin{tabular}{p{0.25\linewidth} | p{0.18\linewidth}| p{0.45\linewidth}}
        Name & Name in package & Description \\
        \hline 
        Local Sidereal Time& lst & Time, in units of the local sidereal time \\
        Transverse Seeing & pt\_seeing & Blurring of the image due to optical distortion by the atmosphere\\
        Band Seeing & band\_seeing & Blurring of the image due to optical distortion by instrumentation\\
        Seeing & fwhm & Combined impact of transverse and band seeing \\
        Airmass & airmass & Line integral of air density along the angle the telescope is observing \\
        Hour Angle & ha & Angular distance between the zenith and the site of interest\\
        Moon Hour Angle & moon\_ha & Angular distance between the zenith and the moon\\
        Moon Elongation & moon\_elongation& Position of the moon in its orbit around the earth\\
        Moon Phase & moon\_phase& Moon phase in a \% of a full moon \\
        Moon Separation & moon\_separation & Angular distance between the target observation location and the moon\\
        Moon Right Ascension & moon\_ra& Right ascension of the moon's position \\
        Moon Declination & moon\_decl & Declination of the moon's position\\
        Moon Airmass & moon\_airmass & Airmass if the telescope was to directly observe the moon\\
        Azimuthal Angle & az & Rotation of the telescope about its axis\\
        Altitude & alt & Height of the apex of the telescope with respect to the ground\\
        Hour Angle & ha & Angular distance between the zenith \\
        Sun Hour Angle & sun\_ha & Angular distance between the zenith and the sun\\
        Sun Airmass & sun\_airmass&  Airmass if the telescope was to directly observe the sun \\
        Sun Right Ascension & sun\_ra& The right ascension of the sun's position \\
        Sun Declination & sun\_decl& The declination of the sun's position\\
        Atmospheric transmission ($\eta$) & sky\_magnitude & The fraction of light from astronomical sources that makes it through the atmosphere~\cite{neilsen_dark_2019} \\
        $\tau$ &tau& Instantaneous measure of sky quality with respect to an observation \\
        $\tau_{eff}$&teff& Measure of observation quality as defined in Equation~\ref{eq:teff}\\

     \end{tabular}
     \label{tab:vars}
 \end{table}

\begin{figure}[h]
    \centering
    \includegraphics[width=.35\textwidth]{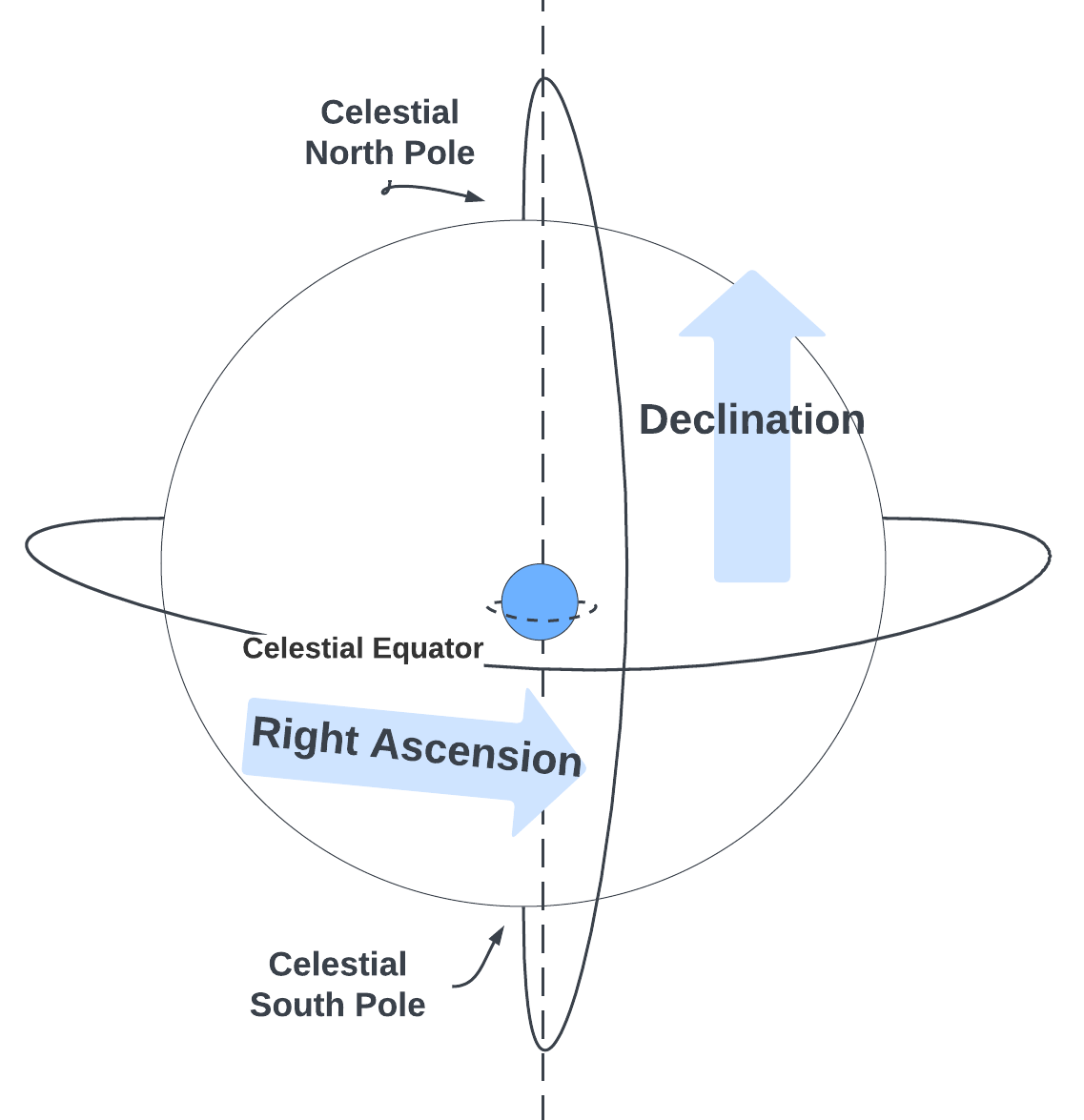}
    \caption{A simple map of the relationship between Right Ascension and Declination. Declination is the angle displaced vertically from the celestial equator, where positive declination denotes an observation in the northern sky, and negative declination denotes the southern sky. Right Ascension is the displacement from the point in the celestial equator where the sun is centered during the Spring equinox.}
    \label{fig:radecl_map}
\end{figure}

\begin{figure}[h]
\centering
\begin{subfigure}[b]{0.4\textwidth}
         \centering
         \includegraphics[width=\linewidth]{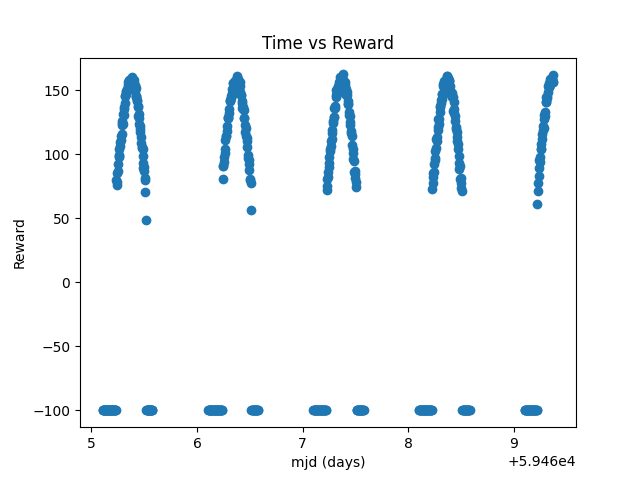}
    \label{fig:reward}
    \caption{}
    \end{subfigure}%
    \hfill
    \begin{subfigure}[b]{0.4\textwidth}
      \centering
      \includegraphics[width=\linewidth]{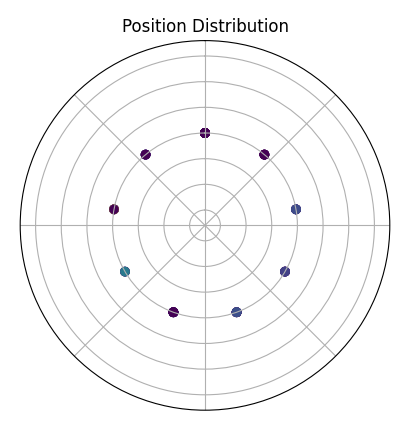}
      \label{fig:position}
        \caption{}
    \end{subfigure}
    \caption{
    An example solution to the problem described in \ref{sec:problem_1}, using a Reinforcement Learning algorithm limited to an equatorial survey ($decl=0$). (a) The values of Equation ~\ref{eq:teff} across multiple nights. The negative values are associated with a negative "penalty" term applied to steps that are outside the user defined allowed range of $\tau_{eff}$. This occurs at dawn and dusk when the sky brightness has dramatically increased. (b) The location of the selected observations projected onto a polar map where the angular distance from vertical ($ra=0$) represents Right Ascension. The lighter colors on sites chosen by the algorithm represents relatively higher reward.}
    \label{fig:rl_solution}
\end{figure}

\clearpage

\bibliographystyle{plainnat}

\end{document}